\documentclass[final,3p,times,twocolumn]{elsarticle}

\usepackage{amssymb}
\usepackage{amsmath}
\usepackage{hyperref}

\usepackage{listings}
\usepackage[dvipsnames,table,xcdraw]{xcolor}
\definecolor{backgroundColour}{rgb}{0.98,0.98,0.98}
\lstdefinestyle{CStyle}{
    backgroundcolor=\color{backgroundColour},   
    commentstyle=\color{ForestGreen},
    keywordstyle=\color{magenta},
    numberstyle=\tiny\color{darkgray},
    stringstyle=\color{Bittersweet},
    basicstyle=\footnotesize,
    breakatwhitespace=false,         
    breaklines=true,                 
    captionpos=b,                    
    keepspaces=true,                 
    numbers=left,                    
    numbersep=5pt,                  
    showspaces=false,                
    showstringspaces=false,
    showtabs=false,                  
    tabsize=2,
    language=C++,
    keywordstyle = [2]{\color{gray}},
    keywordstyle = [3]{\color{RedViolet}},
    keywordstyle = [4]{\color{teal}},
    otherkeywords = {;,<<,>>,++},
    morekeywords = [2]{;},
    morekeywords = [3]{<<, >>,|,=,==,->,&},
    morekeywords = [4]{++}
}

\journal{Computer Physics Communication}

\begin{document}

\begin{frontmatter}

\title{DataPix4: A C++ framework for \\Timepix4 configuration and read-out}

\author[a,b]{V.~Cavallini\corref{cor1}}
\ead{viola.cavallini@unife.it}
\author[b]{N.~V.~Biesuz}
\author[a,b]{R.~Bolzonella}
\author[b]{E.~Calore}
\author[a,b]{M.~Fiorini}
\author[b]{A.~Gianoli}
\author[c]{X.~Llopart Cudie}
\author[a,b]{S.~F.~Schifano}

\cortext[cor1]{Corresponding author}

\affiliation[a]{organization={Università degli Studi di Ferrara},
            addressline={Via Saragat 1},
            city={Ferrara},
            postcode={44122},
            state={Italy},
            country={}}

\affiliation[b]{organization={INFN Ferrara},
            addressline={Via Saragat 1},
            city={Ferrara},
            postcode={44122},
            state={Italy},
            country={}}

\affiliation[c]{organization={CERN},
            addressline={Esplanade des Particules 1},
            city={Geneve},
            postcode={},
            state={Switzerland},
            country={}}

\begin{abstract}
DataPix4 (Data Acquisition for Timepix4 Applications) is a new C++ framework for the management of Timepix4 ASIC, a multi-purpose hybrid pixel detector designed at CERN. Timepix4 consists of a matrix of 448×512 pixels that can be connected to several types of sensors, to obtain a pixelated detector suitable for different applications.
DataPix4 can be used both for the full configuration of Timepix4 and its control board, and for the data read-out via \textit{slow control} or \textit{fast links}. Furthermore, it has a flexible architecture that allows for changes in the hardware, making it easy to adjust the framework to custom setups and exploit all  classes with minimal modification.
\end{abstract}


\end{frontmatter}

\section{Introduction}
\label{sec_introduction}

Timepix4 ASIC is a new hybrid pixel detector designed at CERN in 2019 by the Medipix Collaboration \cite{llopart2022}. It consists of a matrix of 448×512 square pixels with a pitch of 55 $\mu$m and can be connected to different types of sensors, according to the need. Furthermore, thanks to its innovative design, it can be tiled on all 4 sides, allowing the development of detectors with large active area fraction.
Its characteristics make it a multi-purpose ASIC that can be used in a great variety of applications: from fundamental and particle physics to medical applications, biology, archaeological science, material science and more.

An innovative ASIC like Timepix4 must have a strong software counterpart, capable of an optimal management of the hardware system, efficient in its configuration, optimized for its read-out and easy to operate for the final user. Also, the framework must be flexible enough, so that it can be easily adapted to different hardware setups, allowing Timepix4 to be used in different fields without any software modification.

This paper presents a C++ framework designed and implemented ad-hoc for these purposes.
The paper is organized in five main sections: in the first two, a brief description of the Timepix4 ASIC (section \ref{sec_tpx4ASIC}) and the hardware setup architecture (section \ref{sec_hwSetup}) is presented. Then, DataPix4 is introduced, first describing a general overview (section \ref{sec_swArchitecture}), then explaining more in detail the main two parts of the software: configuration of Timepix4 and hardware setup (section \ref{sec_configuration}) and data acquisition (section \ref{sec_dataAcquisition}). The data acquisition section, in particular, will contain three sub-sections describing how to exploit the slow control connection to perform a slow read-out (sub-section \ref{subsecSlowRO}) and the fast links connection for the fast read-out (sub-section \ref{subSec_fastRO}), concluding with a sub-section on the online analysis (sub-section \ref{subsec_online_analysis}), describing the possibility to have an online monitor during the acquisition (paragraph \ref{subsubsec_onlineMon}) and an online clustering (paragraph \ref{subsubsec_onlineClu}). Finally, in section \ref{sec_conclusion} we draw our conclusions.

\section{Timepix4 ASIC}
\label{sec_tpx4ASIC}
\begin{figure}[t]
\begin{center}
\includegraphics[scale=0.35]{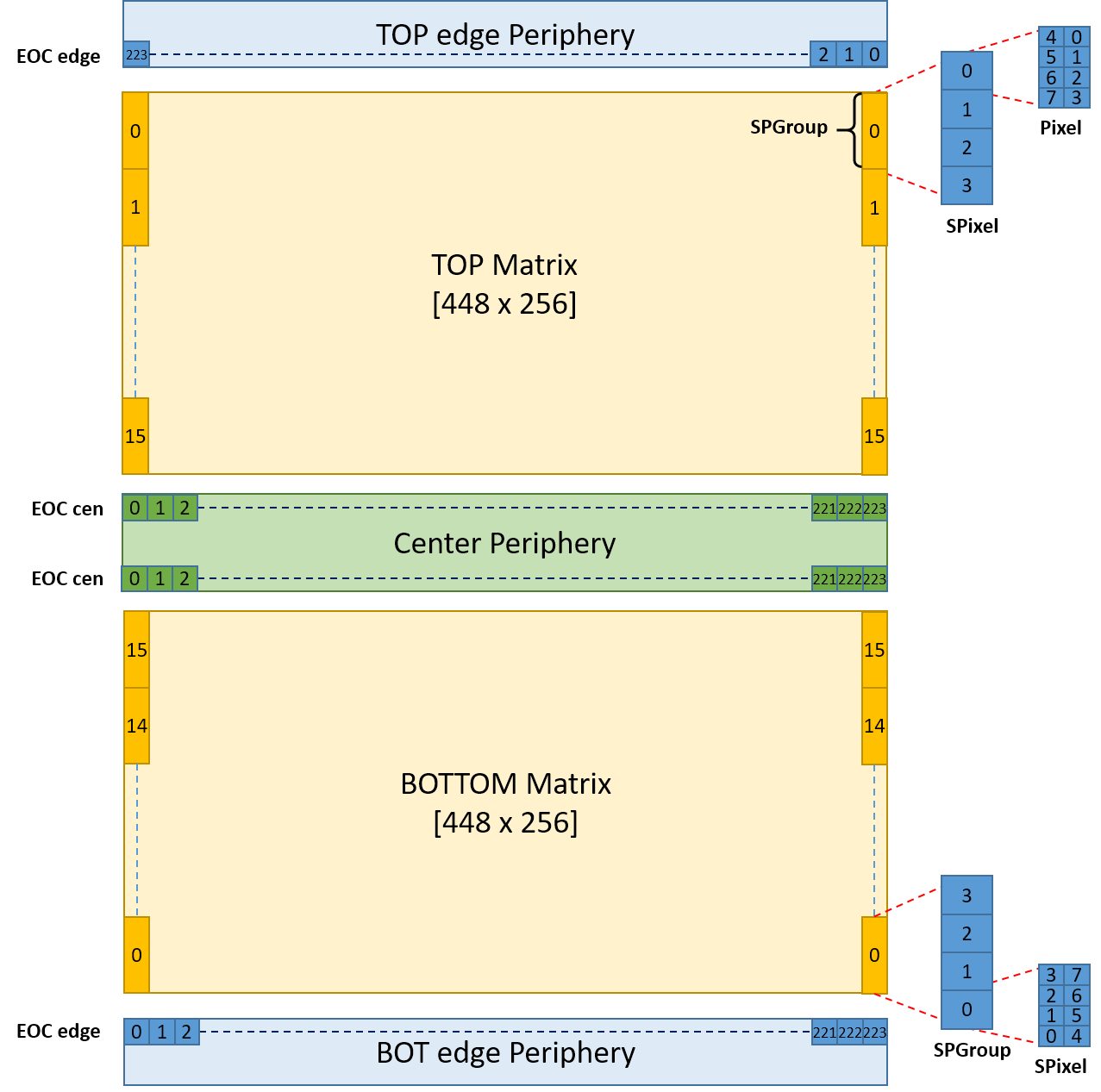}
\end{center}
\caption{Pixel configuration address map, with respect of the Timepix4 matrix organization. Pixels are organized in two half matrices and three peripheries, with different electronic components, are present. In the real design, the center periphery is places under the matrices, so no gap is present.}
\label{fig:ChipPixelOrganization}
\end{figure}

The Timepix4 ASIC has a pixel matrix made by $\sim\thinspace230k$ pixels with a pitch of 55$\thinspace\mu$m, for a total active area of $\sim 7\thinspace$cm$^2$. It can be divided into two sub-matrices, top and bottom, as shown in Fig.~\ref{fig:ChipPixelOrganization}. Those are identical but independent, so each one has its own registers for the configuration, and each one sends the output data through an independent physical channel.

Next to the matrices there are two edge peripheries - TOP and BOTTOM - that are mostly used for read-out and contains electronic components like serializers, routers and End-of-Columns blocks. Between the matrices there's a Center periphery that contains electronic components like the edges ones, bust it also hosts sensors, like temperature and power supply ones, and analog blocks like Analog-to-Digital-Converters (ADC) and Digital-to-Analog-Converter (DAC). In the real design, the Center periphery is placed under the matrices, so no gap is present between top and bottom matrix. A more detailed description of the Timepix4 ASIC can be found at \cite{llopart2022}.

Timepix4 can operate in 2 different modes: \textit{data-driven} or \textit{frame-based}. 
\textit{Frame-based} is a non zero-suppressed read-out mode that periodically reads out all the pixel matrix sending out the photon counting events information collected during a fixed time interval, called \textit{frame}, at a maximum rate of 5 Ghits/mm$^2$/s. Each pixel has two counters (for continuous read/write operation) with programmable depth (8 or 16 bits).
\textit{Data-driven} is, instead, a zero-suppressed, event-driven mode, where Timepix4 generates one data packet every time a pixel has a hit above the configurable threshold. The packet can contain only the coordinate of the pixel and a photon couter with a depth of 24 bits (photon counting mode), or it can contain  more detailed information about the event, such as pixel coordinates, time of arrival, time over threshold (ToA-ToT mode). In \textit{data-driven}, the ASIC can handle a maximum rate of 3.6 Mhits/mm$^2$/s.

\section{Hardware setup}
\label{sec_hwSetup}
\begin{figure}[t]
\begin{center}
\includegraphics[scale=0.25]{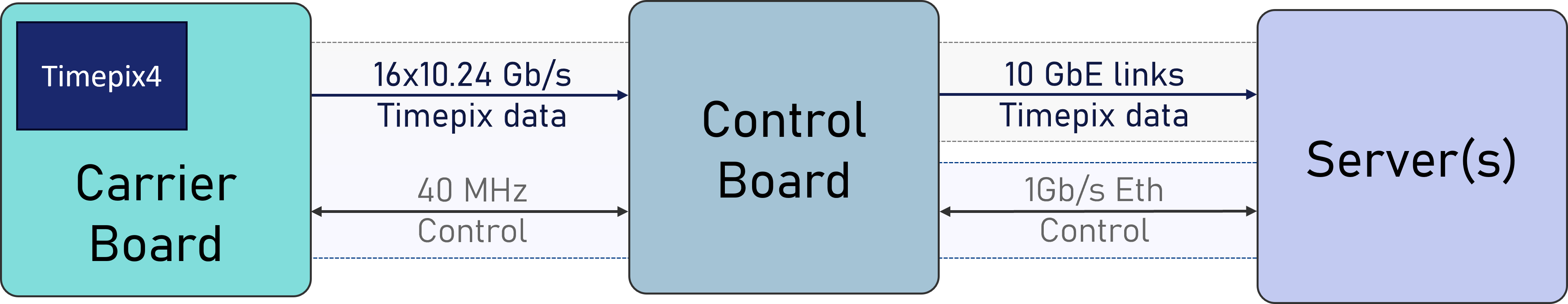}
\end{center}
\caption{Simple scheme of the data acquisition system use to manage Timepix4. 
The ASIC is accomodated on the Carrier Board, that is connected through two type of links to the Control Board. This is then connected to one or more servers through 16 \textit{fast links} at 10 Gb/s and one \textit{slow control} link at 1 Gb/s. Communication via \textit{slow control} is bidirectional, while \textit{fast links} only transmit packets from Control Board to servers.
}
\label{fig:daq_scheme}
\end{figure}

Timepix4 is configured and read out through a data acquisition system composed by a carrier board and a control baord. The control board acts like an intermediary between the carrier baord and the computer, while the carrier board accommodated Timepix4 and provides all the necessary connection between it and the control board. A simple scheme is shown in Fig.~\ref{fig:daq_scheme}.

In the first setup used to test DataPix4, Timepix4 is mounted on a custom chipboard developed by NIKHEF, which provides the connection with the SPIDR4 control board. 
SPIDR4 (Speedy PIxel Detector Read-out 4) is a new read-out system developed by NIKHEF \cite{spidr_website} for communication with Timepix4 ASIC. It acts as an intermediary between Timepix4 and the computer. The chipboard can be directly connected to the control board or a flexible connector may be used to extend the link.

\begin{figure}[t]
\begin{center}
\includegraphics[scale=0.35]{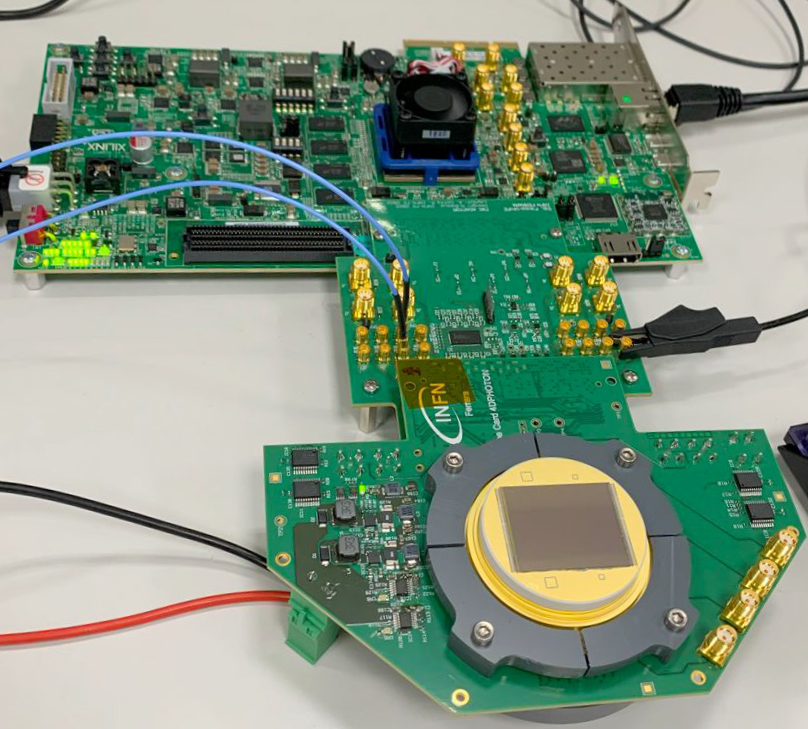}
\end{center}
\caption{Timepix4 mounted on a ceramic PCB connected to the control board developed at INFN Ferrara.}
\label{fig:idaq}
\end{figure}

DataPix4 has also been tested with a different data acquisition system developed at INFN Ferrara \cite{biesuz_PM} and shown in Fig.~\ref{fig:idaq}. This data acquisition system is based on a commercial AMD/Xilinx KCU105 Development Kit, housing a Xilinx Kintex Ultrascale XCKU040-2FFVA1156E FPGA.
This data acquisition system is being developed in the framework of the 4DPHOTON project, whose goal is the production of an innovative photodetector based on a vacuum tube, a MicroChannel Plate and a Timepix4 ASIC as a pixelated anode \cite{fiorini2018}.

\section{DataPix4 architecture}
\label{sec_swArchitecture}
The requirements for the software are high speed and flexibility, followed by user-friendliness. For these reasons, when designing DataPix4, we decided to use an object-oriented paradigm. Exploiting the objects to map the hardware architecture as much as possible, we created different classes for different hardware components, for example the control board or the communication links. In this way, we created a solid structure of classes and objects that is, however, flexible enough to adapt to changes in the hardware. 
The software is completely written in C++, a programming language considerably fast and sufficiently low-level to allow a fast management of data and a complete management of the setup components and the servers memory \cite{CARY199720}.

\begin{figure*}[t]
\begin{center}
\includegraphics[scale=0.60]{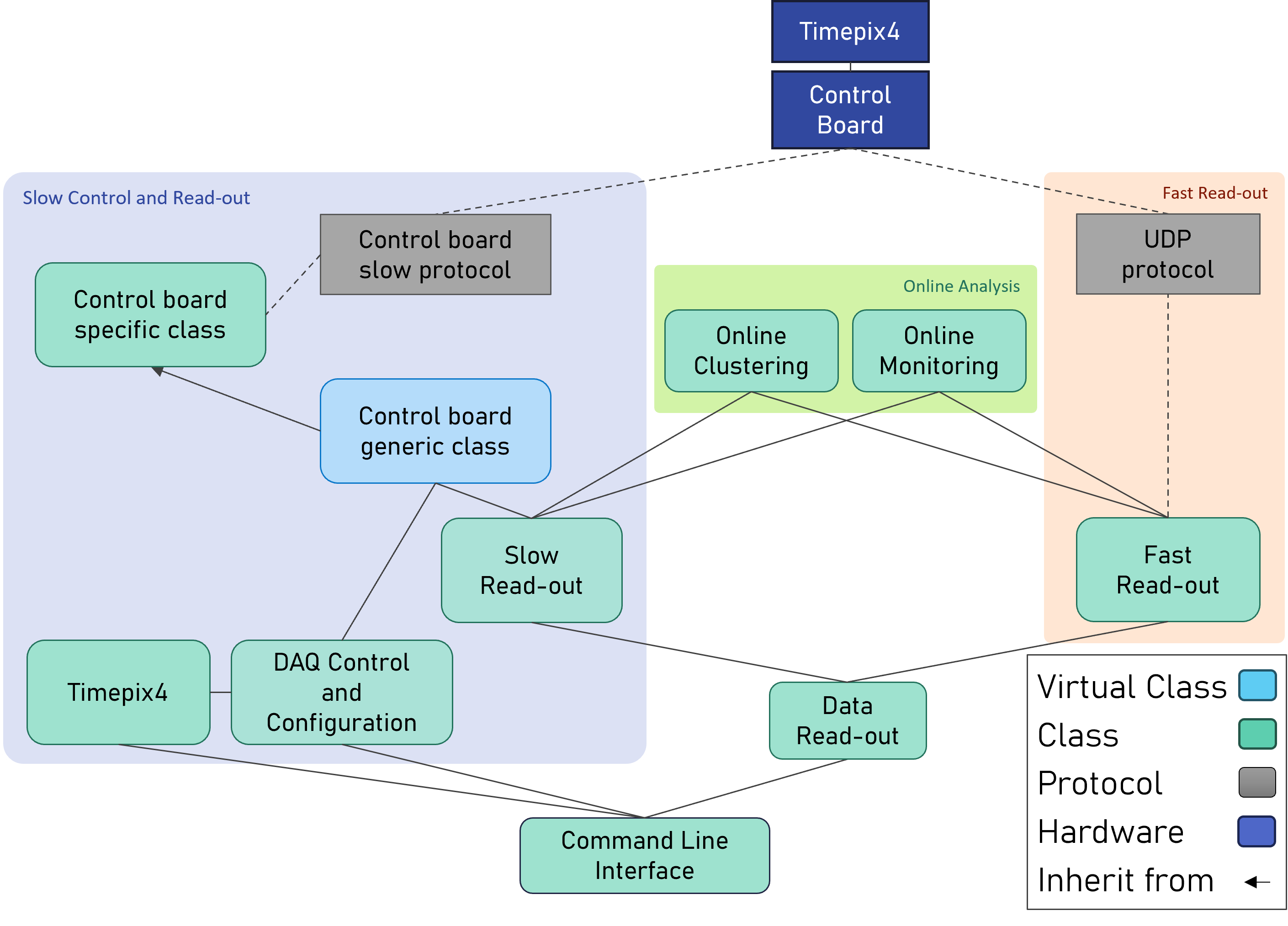}
\end{center}
\caption{DataPix4 classes scheme. The three main parts of the software are visible, in particular: the \texttt{Slow Control and Read-out} manages the configuration and the slow read-out, the \texttt{Fast Read-out} manages the communication via \textit{fast links} and the \texttt{Online Analysis} manages the online visualization and clustering of the events.}
\label{fig:software_scheme}
\end{figure*}

As show in Fig.~\ref{fig:software_scheme}, DataPix4 appears divided in three main parts: the configuration and slow read-out, on the left, the fast read-out, on the right, and the online analysis, in the middle. 
All the classes in the first one manage the slow control operations, in particular the configuration of both Timepix4 and the control board and the slow read-out of the output data. The communication is done using the control board protocol with a client-server paradigm, where the server runs on the board to accept the user's requests.
Since Timepix4 fast links use UDP for communication, the fast read-out classes implements this protocol and therefore they can be used with any Timepix4 setup: the UDP data packets are sent from the Timepix4 chipboard directly to the server, bypassing the control board.

To design a software that can adapt to any custom control board, we exploited the inheritance of C++, creating a virtual \texttt{DAQControl} class that must be implemented in a custom control board-specific class. This must provide the four fundamental methods for the communication: one to perform a reading of a control board register, one to write a value into it, and two for the same operations but on Timepix4 registers. All the other methods of the virtual \texttt{DAQControl} class rely on those four, allowing to have a complete and functional software for every custom setup, once the DAQ derived class is implemented. 

Thanks to the flexible object-oriented design, it is possible to create other custom classes and add them to the main software, in order to personalize it, depending on the needs.

DataPix4 is compiled and tested under Ubuntu 20.04 LTS and Ubuntu 24.04 LTS, but can be compiled and used also under other Linux distribution, depending on the pre-requisite of the control board communication protocol. It was compiled using g++ version 9.4.0 and 13.3.0, but can be compiled with every compiler that supports C++ 17. The installation relies on \texttt{cmake} and \texttt{make}.

To use DataPix4 with custom control board it is necessary to install and correctly configure the control board communication protocol, adding to the \texttt{CMakeLists.txt} file all the instructions needed to compile it and to correctly link its libraries to the DAQ derived class.

\section{Configuration of Timepix4 and Control board} \label{sec_configuration}

The configuration of both Timepix4 and the control board is done via slow control link, using the four fundamental methods: \texttt{readTpx4()}, and \texttt{writeTpx4()}, to perform operations on Timepix4 registers, and \texttt{readBoard()} and \texttt{writeBoard()}, to manage the control board's ones.
Those four functions are implemented in the control board-specific class, since their functioning is strictly dependent on the communication protocol used by the board. This class is the only connection between the control board and the rest of the software\footnote{The \texttt{FastReadout} class just keeps listening on a port, but cannot send messages to the control board.}, as shown in Fig.~\ref{fig:software_scheme}, so it is the only one that needs to be modified if the control board is substituted.

To communicate with the entire setup, the user needs to create an object of the control board-specific class.
\begin{lstlisting}[frame=single, language=C++,style=Cstyle]
  DAQ* board = new DAQ(ip_address, port);
\end{lstlisting}
If the board supports more than one Timepix4, the object constructor, or another related function, may assign to them an ID and/or an index. Those are required arguments for all the Timepix4-related functions, and should be managed by the \texttt{readTpx4()} and \texttt{writeTpx4()} methods, to send requests to the correct Timepix4. If just one ASIC is supported by the board, or if there's no need to use IDs or indexes, the value '0' can be used. 
\begin{lstlisting}[frame=single, language=C++,style=Cstyle]
DAQ* board = new DAQ(ip_address, port);
// Add one Timepix4 to the board object
int tpx_id = board->addTpx(1); 
\end{lstlisting}

Once the \texttt{DAQ} object is instantiated, it can be used to communicate with Timepix4. To write a value on a Timepix4 register, the \texttt{writeTpx4()} function should be used, passing as arguments the address of register, its length and the value to write.
The \texttt{writeTpx4()} function takes the value to write on Timepix4 register as an array of unsigned char. This is necessary to deal also with long registers (up to 512 bytes), given that there are no C++ standard data types longer than 64 bits\footnote{https://en.cppreference.com/w/cpp/language/types }. 
\begin{lstlisting}[frame=single, language=C++,style=Cstyle]
unsigned char *value = (unsigned char *)malloc(2 * sizeof(char));
value[0] = 0x0001;
value[1] = 0x1e00;
int err = board->writeTpx4(tpx_id, idx, reg_address, value, 16);
if (!err) {
    std::cout << Correctly written on Timepix4 reg: << reg_address << std::endl;
}
\end{lstlisting}
However, since the majority of registers has 32 bits or less, a simplified version of this function, taking an unsigned 32-bit integer as argument, has been created.
\begin{lstlisting}[frame=single, language=C++,style=Cstyle]
uint32_t value = 0x00011e00;
int err = board->simpleWriteTpx4(tpx_id, idx, reg_address, value);
\end{lstlisting}

On the opposite, to read back values from Timepix4 registers, the \texttt{readTpx4()} function may be called. This takes the address of the register and its length as arguments, and returns a pointer to an object derived from \texttt{ctrlBoardReplyType}.
\begin{lstlisting}[frame=single, language=C++,style=Cstyle]
ctrlBoardReplyType* reply = board->readTpx4(tpx_id, idx, reg_address, len);
\end{lstlisting}
\texttt{ctrlBoardReplyType} is a virtual class defined in \texttt{DAQ} library that is used to temporarily store the value read from a register into a custom variable. This is necessary because different communication protocols in the control boards may return different data types: for example SPIDR4 board returns a string when reading a register, while other control boards may return an array of unsigned char or unsigned integer. When using a custom control board, it is necessary to create a \texttt{ctrlBoardReplyType} derived class, that implements the following 5 methods, to convert the data read from slow control to 5 different C++ types: \texttt{toUint16()}, \texttt{toUint32()}, \texttt{toInt32()}, \texttt{toFourUInt64()}\footnote{This method returns a vector of uint64\_t. It is useful when reading acquisition data from slow control 256-bit registers, since there are stored 4 x 64-bit data packets.} and \texttt{toString()}.
When calling \texttt{readTpx4()}, the user can directly convert the value into a standard type, making the \texttt{ctrlBoardReplyType} class transparent.
\begin{lstlisting}[frame=single, language=C++,style=Cstyle]
uint16_t reply = board->readTpx4(tpx_id, idx, reg_address, len)->toUint16();
std::cout << "Data read: " << reply << std::endl;
\end{lstlisting}

If DataPix4 is managed by an expert user, the entire setup can be configured manually, writing the correct bytes to all the registers. 

In order to be more user-friendly and to be shareable with user with different expertise levels, higher-level functions have been implemented. Those can semi-automatically configure one or more Timepix4 settings, writing the dedicated registers, so the user does not need to manually configure the entire setup register-by-register.
The functions are written in the \texttt{DAQControl} virtual class, so that it is possible to access them with every DAQ-derived object.
To fully configure Timepix4 in a fast and secure way, the user can initially call \texttt{confThreshold()} and \texttt{confTimepix4()} methods. 
While the first one configures the Timepix4 DACs that set the global matrix threshold, the second one performs a full Timepix4 basic configuration.
To set the most important configuration parameters, the user needs to fill out a configuration file, as shown in the example.
This file is processed by \texttt{Timepix4::createTpxConfStructFromFile()}, a function that returns a C++ struct with all the parameters. The struct can be exploited by the user, to get or change some of the configuration parameters, and has to be passed to the two higher-level configuration functions.
\begin{lstlisting}[frame=single, language=C++,style=Cstyle]
int POWER_MODE=0
bool POLARITY=1
// Sending out control packet?
bool ENABLE_STATUS_PACKET=1
// Sending out heartbeat packet?
bool HEARTBEAT_ENABLE=0      
bool ENABLE_TOA=1           
bool ENABLE_TDC=1
bool ENABLE_DLL_COLS=1
// Are you using fast link?
int USING_FAST_LINKS=0       
// Read data via slow control?
bool READ_OUT_SLOW_CONTROL=1 
// Is test-pulse on?
bool TP_ON=1                 
// How many test-pulses?
int NUM_TP=2000              
// Time a single test-pulse is on
float TP_TIME_ON=0.0001      
// Time between 2 test-pulses
float TP_TIME_OFF=0.0005     
// How many s the shutter is open
float SHUTTER_TIME_ON=0.005  
// How many s the shutter is closed
float SHUTTER_TIME_OFF=0.001 
// Data-driven or frame-based?
int READOUT_MODE=1           
[ ... ]
\end{lstlisting}

Once \texttt{confTimepix4()} and \texttt{confThreshold()} functions are called, Timepix4 is almost ready to start working using the mode and the parameters specified in the configuration file. Those functions are particularly useful both for the first tests, to check the correct behaviour of the setup, and for an initial configuration at the beginning of every script, that can then be fine-tuned using specific APIs or writing Timepix4 registers, according to the needs.
An example of Timepix4 configuration is shown in the box below.
\begin{lstlisting}[frame=single, language=C++,style=Cstyle]
// Create a configuration struct from the configuration file
Tpx4Config confStruct=Tpx4::createTpxConfStructFromFile("tpx4.conf");
// Use the struct to configure Timepix4
board->confTimepix4(tpx_id, idx, confStruct);
// Use values from the configuration struct
int THR_e = confStruct.Values["THRESHOLD"];
std::cout << "Using threshold: " << THR_e << std::endl;
// Create a threshold configuration struct
Tpx4ThresholdConf thrStruct = Tpx4::createTpxThrConfFromStruct(confStruct, linearize_thr, thr, fbk);
// Change struct's values at run-time
thrStruct.verbose = true;
// Configure Timepix4 thresholds using the struct
board->confThreshold(tpx_id, idx, thrStruct, BOT);
\end{lstlisting}

Before starting the acquisition, the user needs to configure also Timepix4 pixels matrix. Each of the $\sim$230$\thinspace$k pixel has 8 dedicated bits that may be used to set the local DACs used to tune the local threshold value (5 bits) and whether or not enable its power (1 bit), test-pulse (1 bit) and mask (1 bit). The \texttt{pixelConf()} method can be called to convert the four parameters into a 8-bit unsigned value with the codified information. Timepix4 indexes its pixels with a [double column, super pixel, pixel] coordinate system, as shown in Fig.~\ref{fig:ChipPixelOrganization}, and has 448 4096-bit registers to configure all the pixels, one for each double column of the matrix. 
To make the configuration more user-friendly, a dedicated function has been implemented, \texttt{configPixel()}. This takes 2 arguments: the first is a 3D vector (corresponding to the 3D coordinate system described above) containing the 8-bit value for each pixel, while the second is the matrix periphery in which the values must be written (choosing from \texttt{top} or \texttt{bottom}). The function performs all the necessary casting and writing. The vector must follow the [double column, super pixel, pixel] scheme, but the user can choose to configure all the pixels with a (X, Y) coordinate system and then use a dedicated function to convert the coordinates, \texttt{addrColRowToEoCSpPix()}, as shown in the example below.

\begin{lstlisting}[frame=single, language=C++,style=Cstyle]
typedef vector<vector<vector<uint8_t>>> tpxConfigMatrix;
vector<int> addr(4);
bool power = true, tp = false, mask = false;

// Create the two pixels vectors
tpxConfigMatrix pixelConfigTop(224, vector<vector<uint8_t>>(16,vector<uint8_t>(32)));
tpxConfigMatrix pixelConfigBot(224, vector<vector<uint8_t>>(16,vector<uint8_t>(32)));

// Load pixels' DAC values form a file
vector<vector<int>> dac_thr_pattern(MATRIX_SIZE_X,vector<int>(MATRIX_SIZE_Y));
Timepix4::fileToMatrix("dac_values.dat", dac_thr_pattern);

// A double for-loop configurates each pixel
for (int x = 0; x < MATRIX_SIZE_X; x++) {
  for (int y = 0; y < MATRIX_SIZE_Y; y++) {
// Convert [X,Y] coordinates into [double column,super pixel,pixel]
    addr = board->addrColRowToEoCSpPix(y, x);
    if (addr[0] == TOP) {
      pixelConfigTop[addr[1]][addr[2]][addr[3]] = board->pixelConf(dac_thr_pattern[x][y], power, tp, mask);
    } else {
      pixelConfigBot[addr[1]][addr[2]][addr[3]] = board->pixelConf(dac_thr_pattern[x][y], power, tp, mask);
    };
  };
};
// Call the API that writes the configurations on the registers
board->ConfigPixel(tpx_id, idx,pixelConfigTop, TOP);
board->ConfigPixel(tpx_id, idx,pixelConfigBot, BOT);
\end{lstlisting}

There are other APIs to change the pixel matrix configuration, for example the \texttt{ConfigAllPixels()}, that takes as arguments the four parameters (DAC value, power, test-pulse and mask) and set every pixel with the same values, or the \texttt{ConfigSuperPixelCol()}, that configure every double column instead of every pixel.

Once both Timepix4 and its pixel matrix are correctly configured, the data acquisition can be started and the user can begin to collect the data packets.

\section{Data acquisition}
\label{sec_dataAcquisition}
Once the setup is correctly configured, there are two ways to acquire the data: slow and fast read-out, each with dedicated and independent threads.

\subsection{Slow Read-out}
\label{subsecSlowRO}
Timepix4 slow control link allows a read-out of the data up to a maximum bandwidth of 1Gb/s. There are two dedicated Timepix4 registers, one for each edge periphery, that, during the acquisition, will contain the data packets. In particular, each register is 256-bit long, so it will contain a maximum of four 64-bit data packets. The reading of those registers is destructive, so the data packets are deleted and replaced with newer ones from a FIFO every time the register is read. If data is not read in time, it will be overwritten with new data. This implies that a thread must continuously poll the register(s) in order to read all output data. 

The \texttt{slowReadout} class constructor takes as input the following values:
\begin{itemize}
\setlength{\itemsep}{1pt}
\item board: DAQ*, a pointer to the control board.
\item tpx$\_$id and idx: int, the Timepix4 ID and index.
\item rawReadout: bool, true if you want output data stored in binary files, false if you want text output files with a uint64$\_$t value for each packet.
\item outputPath: std::string, defines the path in which store the output files.
\item onlineMonitoring\footnote{Online monitoring and online clustering will be described more in detail in section \ref{subsec_online_analysis}\label{fn:analisys_slow}.}: bool, true if you want to run an online monitoring\footnote{This requires ROOT framework installed.} while acquiring data.
\item onlineClustering\footref{fn:analisys_slow}: bool, true if you want to process and clusterize data during the acquisition.
\end{itemize}

The \texttt{slowReadout} object has some \textit{setter} methods\footnote{Every \textit{setter} method has the corresponding \textit{getter}} to configure the data acquisition. In particular:
\begin{itemize}
\setlength{\itemsep}{1pt}
\item \texttt{setPeriphery(Peryphery side)}: specifies the periphery (top, bottom or both) from which you want to read data. Default is 'Peryphery::BOTH'.
\item \texttt{setBufferingLen(int len)}: configures the length of the array that is used to bufferize the writing on file operation. Default is 100 packets.
\item \texttt{setWriteOnFile(bool writing)}:allow or disallow the writing of the read-out data on file\footnote{This flag can be useful when launching test acquisition with online analysis flags. In this way, the user can see the data in real-time and can adjust Timepix4 position (for example in a test-beam) without using disk space and extra time to write data in memory.\label{fn:writeonfile}}.
\end{itemize}

An example of the instantiation of a \texttt{slowReadout} object is shown in the code below.

\begin{lstlisting}[frame=single, language=C++,style=Cstyle]
#include "slowReadout.h"

// Creates one object to acquire data 
slowReadout *slowThread = new slowReadout(&board, tpx_id, idx, rawReadout, outputPath, onlineMonitoring, onlineClustering);
// Set which periphery to read from (BOTH is default)
slowThread->setPeryphery(Periphery::TOP);
\end{lstlisting}

When the object has been created and configured, acquisition can be started.
\begin{lstlisting}[frame=single, language=C++,style=Cstyle]
// Start acquisition
slowThread->startSlowReadout();
\end{lstlisting}

Once acquisition is finished, the stop function may be called.
\begin{lstlisting}[frame=single, language=C++,style=Cstyle]
// Stop acquisition
slowThread->stopSlowReadout();
\end{lstlisting}

When launching the acquisition, a thread for each Timepix4 half matrix is created and begin polling the corresponding register.
The choice of creating one object from each half was made for optimization purposes: instead of calling \texttt{readTpx4()} function in an infinite loop to perform one reading at a time, a \texttt{readTpx4Repeat()} function has been implemented, to read multiple times the same register. This allowed the overhead of a single function call to be spread over multiple reads, instead of using one call for a single register’s read. 

Once the acquisition is completed, the \texttt{stopSlowReadout()} function terminates the dedicated thread(s) and saves the last data on file.

Since it needs to continuously polls a register, the \texttt{slowReadout} class is strictly dependent on the type of control board used and it needs to implement its communication protocol. However, the read-out function relies only on the \texttt{readTpx4()} method, the same also used for the configuration of Timepix4. This implies that, in the event of a control board change, the user does not need to re-implement other methods in addiction to the 4 described in section \ref{sec_configuration}, to have a working slow read-out class.

While re-writing its control board-specific class, the user just needs to be careful about concurrence. The slow read-out thread has been designed to exploit the \texttt{readTpx4()} function, to make it easier and faster to adapt to every control board. This choice, however, implies that the \texttt{readTpx4()} function may be used from more than one thread at the same time. This is not a problem if the method is atomic, but can create undefined behaviors if there are more read or write operations inside the \texttt{readTpx4()} function\footnote{For example, to perform a single \texttt{readTpx4()}, you may need to write different control board registers to choose the correct Timepix4 register's number, length and offset you want to read, then you may need to read the extracted value from a dedicated control board register. All those operations need to be done consecutively and without any interruption.}. In this case, a mutex or an equivalent synchronizing method needs to be exploited when re-writing \texttt{readTpx4()} method.

\subsection{Fast Read-out}
\label{subSec_fastRO}
\begin{figure}[t]
\begin{center}
\includegraphics[scale=0.40]{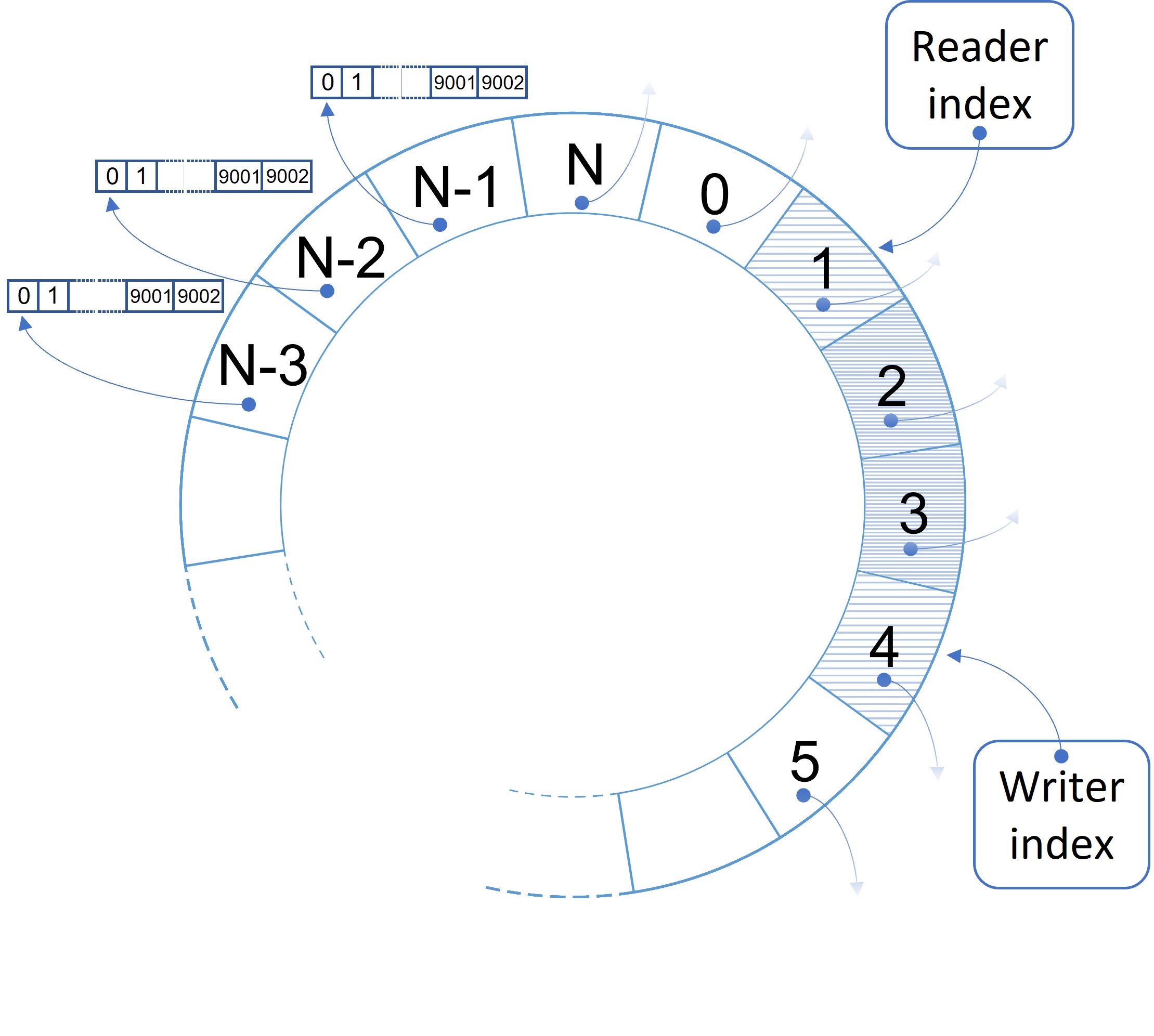}
\end{center}
\caption{Scheme of the circular buffer used for fast read-out. Each element is a pointer to a 9002 bytes long array where a UDP packet is stored. Two mutexes protect the entire structure.}
\label{fig:circular_buffer}
\end{figure}

In addition to the slow control connection, as described in subsection \ref{subsecSlowRO}, Timepix4 also features 8 optic links per half matrix for high-speed readout with a configurable speed, from 40 Mb/s to 10.24 Gb/s per link, allowing a maximum total bandwidth of $\sim$160 Gbps. Thus, a dedicated class has been implemented, called \texttt{fastReadout}. Like the other read-out class, it contains two main methods: \texttt{startFastReadout()} and \texttt{stopFastReadout()}, to manage the data acquisition.
Since this read-out method needs to be able to handle a fast data flow, the class uses two independent threads to catch and store the data for each link. The threads run using a reader-writer paradigm on a shared circular array, shown in Fig.~\ref{fig:circular_buffer}. In particular, one thread listens to the UDP port, catches the data and writes it on the shared buffer, while the other reads the data from the buffer and stores it on files. 
To optimize the server’s memory, the shared structure is not one big array of char; it is, instead, a smaller circular array of char pointers, that point to independent heap memory areas. Those little arrays are 9002 8-bit elements long, this is, in fact, the maximum dimension of jumbo frame packets, plus 2 bytes to store the number of data actually received on every packet, between 1 and 9000. There are just 2 mutexes, one for the index that keeps track of the element currently read and protects the small array while the reader thread is accessing it, and one for the writer index, to protect the access to the small array while the writing thread is handling it.
Since every time a thread interacts with the structure, it reads or writes just one of the small arrays without altering the main buffer, there’s no need to use other mutexes to protect the entire structure.

The \texttt{fastReadout} object automatically connects to one or more UDP ports and keeps listening to the data that arrives on that ports. 
The \texttt{fastReadout} object constructor takes the following arguments:
\begin{itemize}
\setlength{\itemsep}{1pt}
\item IPs: std::vector$<$std::string$>$, vector containing the IP addresses to bind (one for each link).
\item ports: std::vector$<$int$>$, vector containing the ports' numbers to bind (one for each link).
\item outputPath: std::string, defines the path in which store the output files.
\item onlineMonitoring\footnote{Online monitoring and online clustering will be described more in detail in section \ref{subsec_online_analysis}\label{fn:analisys_fast}}: bool, true if you want to run an online monitoring\footnote{This requires ROOT framework installed} while acquiring data.
\item onlineClustering\footref{fn:analisys_fast}: bool, true if you want to process and clusterize data during the acquisition.
\end{itemize}

The \texttt{fastReadout} object also has some \textit{setter} methods\footnote{Every \textit{setter} method has the corresponding \textit{getter}.} to further configure the data acquisition. In particular:
\begin{itemize}
\setlength{\itemsep}{1pt}
\item \texttt{setTimeout$\_$us(int timeout$\_$us)}: set the UPD socket timeout in $\mu$s. Default is 100 $\mu$s.
\item \texttt{setMaxSize(int maxSize)}: set the maximum size (in bytes) of the output file. When this size is reached, the file is closed and a new one is opened. Default is 1 GB.
\item \texttt{setVerbose(bool verbose)}: allow read-out threads to print on terminal every second how many packets are captured and stored and the total amount of data received. Default is true.
\item \texttt{setWriteOnFile(bool write$\_$on$\_$file)}: allow or disallow the writing of the data on file\footref{fn:writeonfile}. Default is true.
\item \texttt{setOutputPath(string outputPath)}: set the path in which store the output data. This is useful to change path in between acquisitions, without re-allocating a read-out object.
\end{itemize}

An example is shown in the code below.

\begin{lstlisting}[frame=single, language=C++,style=Cstyle]
#include "fastReadout.h"

std::vector IPs = {IP1, IP2};
std::vector ports = {port1, port2};
fastReadout *fastThread = new fastReadout(IPs, ports, outputPath, onlineMonitoring, onlineClustering);
// Set socket timeout in us
fastThread->setTimeout_us(500);
// Do not write data on file
fastThread->setWriteOnFile(false);
\end{lstlisting}

Once the objects are instantiated, the fast read-out can be started.
\begin{lstlisting}[frame=single, language=C++,style=Cstyle]
// Start acquisition
fastThread->startFastReadout();
\end{lstlisting}

At the end of the acquisition, to correctly join all the threads, the stop function should be called.
\begin{lstlisting}[frame=single, language=C++,style=Cstyle]
// Stop acquisition
fastThread->stopFastReadout();
\end{lstlisting}

As long as the received packets use UDP protocol and do not exceed the maximum size of 9000 bytes, they are accepted and stored correctly. This makes the control board transparent to the entire \texttt{fastReadout} class, allowing for changes in the hardware with no need to re-implement those methods.
Moreover, the computer used for fast acquisition may be different from the one used for the configuration.

\subsection{Online analysis}
\label{subsec_online_analysis}
DataPix4 includes two classes for real-time processing of the data: \texttt{onlineClustering} and \texttt{onlineMonitoring}, described in the following sections.
Both the algorithm for clustering and monitoring are slower than the acquisition one, for this reason, both the classes uses C++ independent threads and dedicated memory areas. If the event rate is high, the online analysis may not be able to process all of them and miss a few data packet, but this does not compromise the data acquisition, that will be able to capture all the data, since the file storage is performed by a different, independent thread.

\subsubsection{Online clustering}
\label{subsubsec_onlineClu}
Clustering is made on the basis of spatial and temporal proximity. This means that two hits are grouped in the same cluster if they hit neighboring pixels in Timepix4 matrix and if they hit Timepix4 respectively at time $t_1$ and $t_2$, so that $|t_2-t_1| \leq \Delta t_{hits}$, where $\Delta t_{hits}$ is a parameter set by the user, indicating the maximum time elapsed between to hits belonging to the same cluster.

Every time a new packet arrives, the clustering algorithm searches for all the events that might belong to the same cluster. If one or more events are found, the new event is added to a pre-existing cluster, otherwise it will start a new one. 

When the differences between the times of arrival of the events in a cluster and the time of arrival of the current processed event is grater than a $\Delta T_{clusters}$, the cluster is written on file and the memory is marked as over-writable. This ensures that the RAM memory that is used during the acquisition remains approximately constant, allowing for long data acquisitions. The $\Delta T_{clusters}$ parameter can be configured by the user and it is a lower limit of the time elapsed between two clusters.

The online clustering algorithm saves a ROOT file \cite{ANTCHEVA20092499}. The file contains a \texttt{TTree} structure with the following branches:
\begin{itemize}
\setlength{\itemsep}{1pt}
    \item ClusterID: an integer representing the ID of the cluster.
    \item ClusterSize: number of pixel hits in the cluster.
    \item XCenterOfGravity: the weighted X coordinate average of all pixel hits in the cluster.
    \item YCenterOfGravity: the weighted Y coordinate average of all pixel hits in the cluster.
    \item Charge: the total charge of the cluster.
    \item ToT: the total Time over Threshold of the cluster.
    \item ToACenterOfGravity: the weighted Time-of-Arrival average of all pixel hits in the cluster.
    \item Xs: a vector containing all X coordinates of pixel hits in the cluster.
    \item Ys: a vector containing all Y coordinates of pixel hits in the cluster.
    \item ToAs: a vector containing all Time of Arrivals of pixel hits in the cluster.
    \item ToTs: a vector containing all Time over Threshold of pixel hits in the cluster.
    \item Charges: a vector containing all charges of pixel hits in the cluster.
\end{itemize}
The \texttt{TTree} structure is filled every time a new cluster is closed, so it will have one entry for each cluster.

To use online clustering when acquiring data, the user just needs to set to \textit{true} the corresponding flag, when instantiate the read-out object, as shown in the example below.
\begin{lstlisting}[frame=single, language=C++,style=Cstyle]
bool online_clustering = true;
// For slow read-out
slowReadout *slowThread = new slowReadout(&board, tpx_id, idx, rawReadout,outputPath, online_monitoring, online_clustering);
// For fast read-out
fastReadout *fastThread = new fastReadout(IPs, ports, outputPath, online_monitoring, online_clustering);
\end{lstlisting}

Depending on the online monitoring flag, there are two cases:
\begin{itemize}
    \item Online monitoring is set to \textit{false}: online clustering runs and saves the ROOT files containing the clusterized events information.
    \item Online monitoring is set to \textit{true}: online clustering runs and saves the same file as in the previous case. Moreover, it writes information also in a shared memory, where the online monitoring thread can read them to show the cluster statistics in real-time, as described in the next section.
\end{itemize}

\subsubsection{Online monitoring}
\label{subsubsec_onlineMon}
For both slow and fast read-out, it is possible to have also a real-time online monitoring of the data received. This is performed by a separated thread that communicates with the read-out ones, using a shared memory where the packets are stored. 
To use this functionality, the user just needs to set the corresponding flag when instantiating the read-out object.
\begin{lstlisting}[frame=single, language=C++,style=Cstyle]
bool online_monitoring = true;
// For slow read-out
slowReadout *slowThread = new slowReadout(&board, tpx_id, idx, rawReadout,outputPath, online_monitoring, online_clustering);
// For fast read-out
fastReadout *fastThread = new fastReadout(IPs, ports, outputPath, online_monitoring, online_clustering);
\end{lstlisting}

If the online monitoring flag is set to \textit{true}, a popup window will open up when acquisition starts.
There are two different cases:
\begin{itemize}
    \item Online clustering is set to \textit{false}: in this case, the window will display a hitmap\footnote{The type of graph can be changed depending on the need. The user can choose, for example, to plot an histogram instead.} of Timepix4 matrix, with a color scale related to the pixels photon counting.
    \item Online clustering is set to \textit{true}: in this case, the window will display four different plots also with statistics about the clusters. The first plot shows the same hitmap as in the other case. The second plot is an histogram of Time over Threshold values, one value for each cluster ( $\mathrm{ToT_{cluster}} = \sum \mathrm{ToT_{hit}}$ ). The third one is an histogram of clusters' dimensions (number of hits belonging to them). Lastly, the fourth plot is an histogram of the clusters total calibrated charge.
\end{itemize}

\begin{figure}[t]
\begin{center}
\includegraphics[scale=0.35]{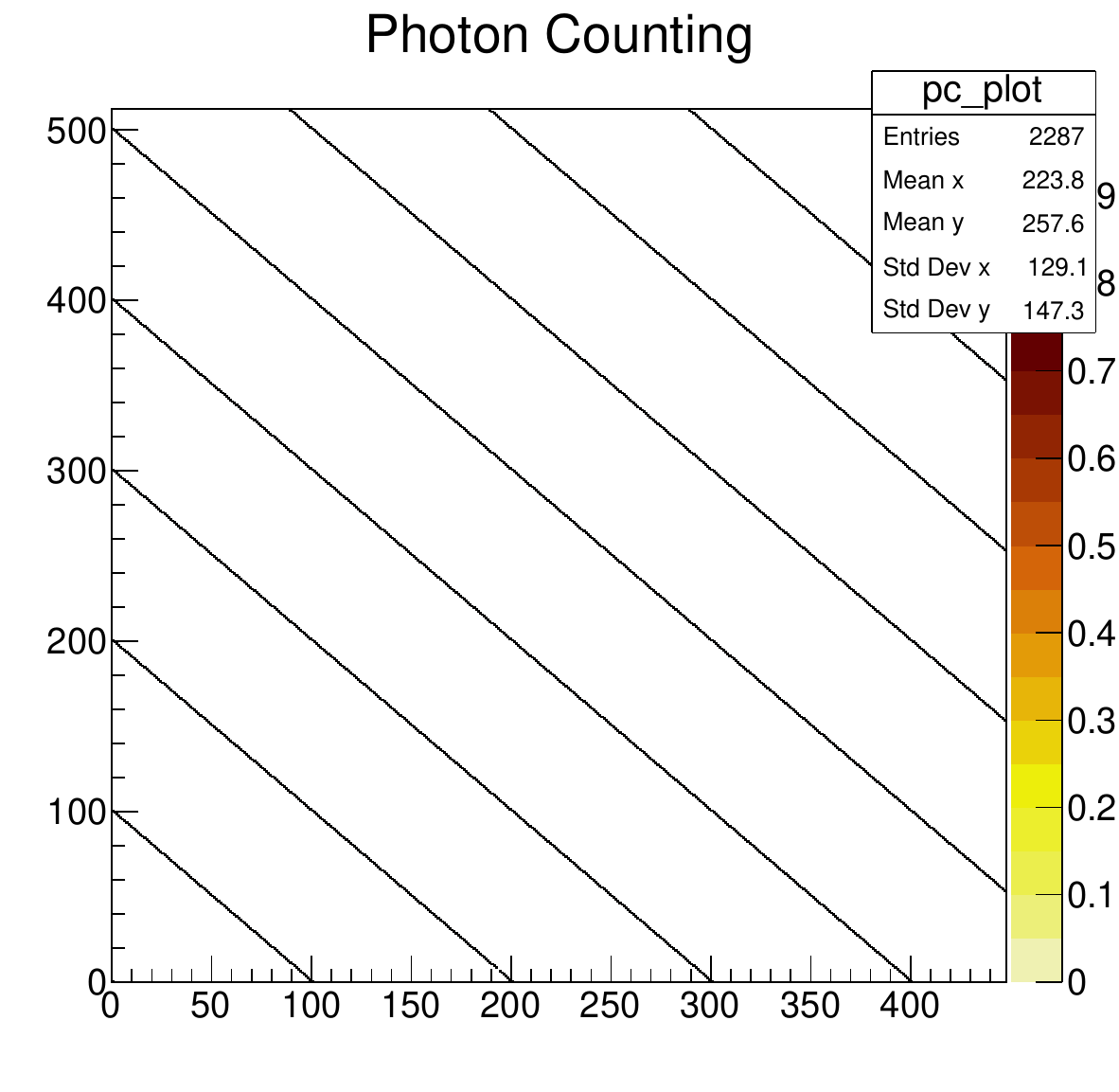}
\end{center}
\caption{Online monitoring hitmap plot when clustering is off. Is this example, only pixels on some diagonals where not masked, forming a pattern. One test pulse was sent to the entire matrix, but only not masked pixels generated a data packet.}
\label{fig:online_mon}
\end{figure}

\begin{figure*}[t]
\begin{center}
\includegraphics[scale=0.60]{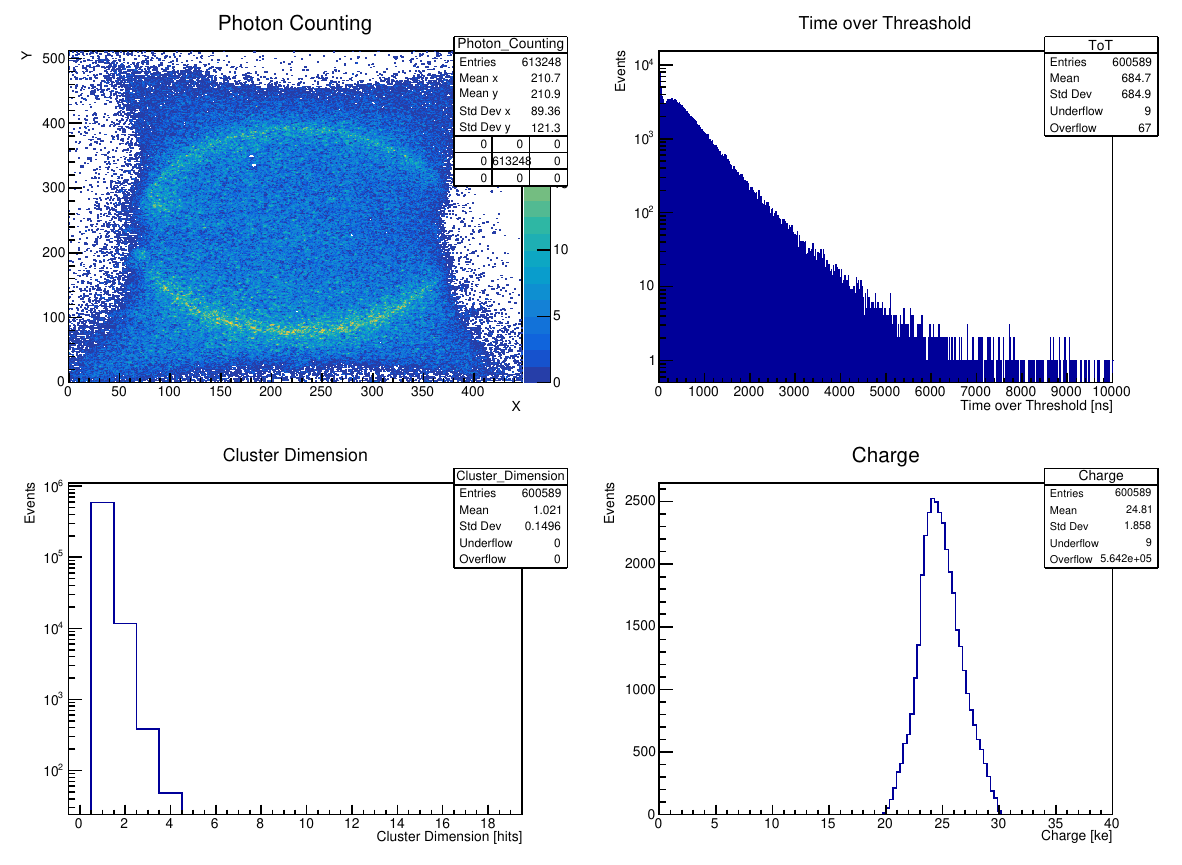}
\end{center}
\caption{Online monitoring plot when clustering is on. This example was taken from an acquisition performed with the 4DPHOTON photodetector during a test beam at CERN. On the upper left, a Cherenkov ring can be seen.}
\label{fig:online_mon_clu}
\end{figure*}

An example of a popup window when clustering if off is shown in Fig.~\ref{fig:online_mon}, while an example with clustering on is shown in Fig.~\ref{fig:online_mon_clu}.
All the graphical part is done using ROOT libraries \cite{BRUN199781}.

Once the acquisition terminates, the online monitoring thread is automatically stopped and the final picture is saved in PDF format.

\section*{Conclusions} \label{sec_conclusion}
A complete and ready-to-use framework for the optimal management of Timepix4 has been developed. DataPix4 can be used both for configuration and for the read-out of the data thanks to the dedicated classes. It can also be expanded with personalized classes for different types of clustering, offline analysis or any other custom algorithm.
Since Timepix4 is a multi-purpose ASIC that can be exploited in different fields and applications, this software has been designed to be flexible and easy to adapt to a new hardware setup. At the moment, it is used with two different hardware setups.

DataPix4 has been widely used at the University of Ferrara, in particular to perform calibrations and timing resolution measurements for the characterization of the Timepix4 ASIC \cite{Bolzonella_2024} and it has been exploited during a characterization of a Timepix4 silicon assembly using monochromatic X-rays at the Elettra synchrotron facility (Trieste, Italy) \cite{feruglio2024}.
Moreover, in the framework of the 4DPHOTON project \cite{fiorini2018}, DataPix4 was used to debug and test the Timepix4 mounted on a tailored ceramic PCB (Fig.~\ref{fig:idaq}) and the first photodetector prototypes equipped with photocathode and microchannel plates. In particular, it was exploited during a test beam at CERN where the first 4DPHOTON photodetector prototype was tested using Cherenkov light.
During both the test beams, almost one Terabyte of data was successfully acquired and stored, while hundreds of Gigabytes of data were processed online and offline, allowing the user to see come of the results of the acquisition in real-time.

\section*{Program availability}
DataPix4 is available exclusively to members of the Medipix4 Collaboration, because it relies on proprietary Timepix4 ASIC intellectual property.

The software has been developed using Baltig as version control tool. In particular, there are two main repositories:
\begin{itemize}
    \item Main software: all the source code that is independent from the control board (for example the fast readout and the online analysis classes) is stored in this repository.
    \item Custom control board classes: contains the source code strictly dependent on the control board (for example the control board specific class). It should contain a \texttt{CMakeLists.txt} file with the instruction to compile and link the board communication protocol. The CMake file should also include the \texttt{CMakeLists.txt} file included in the main software repository. Every control board has a different repository.
\end{itemize}
The control board specific repository should include the main software's one as a sub-module. When cloning the project, the user just need to clone the repository corresponding to his control board with the \texttt{--recursive} option.

\section*{Acknowledgments}
This work was carried out in the context of the Medipix4 Collaboration based at CERN, and in the framework of the MEDIPIX4 project funded by INFN CSN5.
This project has received funding from the European Research Council (ERC) under the European Union's Horizon 2020 research and innovation programme (Grant agreement No. 819627, 4DPHOTON project).

\bibliographystyle{elsarticle-num-names} 
\bibliography{references.bib}

\end{document}